\documentstyle[a4,12pt,lathuile,amsmath,graphics,epsfig,citesort]{article}
\newcommand{\MeV}{\mbox{\rm MeV}}

\begin{document}
\begin{flushright}
{\small
CERN-TH-2002-213\\
IFUM-721/FT\\
FTUAM-02-314
}
\end{flushright}
\begin{center}
{\Large \bf The solar neutrino puzzle: present situation and future 
scenarios}\\[0.2cm]

{\large P.~Aliani$^{a\star}$, 
V.~Antonelli$^{a\star}$,
R. Ferrari$^{a\star}$,
M.~Picariello$^{a\star}$, 
E.~Torrente-Lujan$^{abc\star}$
\\[2mm]
$^a$ {\small\sl Dip. di Fisica, Univ. di Milano},
{\small\sl and INFN Sez.  Milano,  Via Celoria 16, Milano, Italy}\\
$^b$ {\small\sl Dept. Fisica Teorica C-XI, 
Univ. Autonoma de Madrid, 28049 Madrid, Spain,}\\
$^c$ {\small\sl CERN TH-Division, CH-1202 Geneve}\\
}

\end{center}

\baselineskip=14.5pt
\vspace{0.2cm}
\begin{abstract}
We present a short review of the existing 
evidence in favor of neutrino mass
and neutrino oscillations which come from different kinds of experiments.
We focus our attention in particular on solar neutrinos, presenting a
global updated phenomenological analysis of all the available data and
we comment on different possible future scenarios.   

\vspace{5.8cm}

{\it \noindent Expanded version of the contribution to 
appear in the Proceedings of ``Les Rencontres de 
Physique de la Vallee d'Aoste'', February 2002.} 
\end{abstract}
\baselineskip=17pt

\vfill
{\small \noindent 
$\star$ email: paul@lcm.mi.infn.it, vito.antonelli@mi.infn.it,
 marco.picariello@mi.infn.it, torrente@cern.ch}

\vphantom{\vspace{3cm}}
\newpage

\section{Introduction}
Seventy years after Pauli's \cite{Pauli} 
proposal of its existence and almost 
half a century after its discovery~\cite{CowanReines}, the neutrino still 
plays a central role 
in elementary particle physics. The main problem of determining 
whether it is 
a massive or massless particle seems to have been solved 
after the last evidences, 
coming mainly from the solar and atmospheric neutrino experiments, 
but we still 
have to answer important questions. We don't know, 
for instance, whether it is 
a Majorana or Dirac particle, nor have we a unique natural 
explanation of its lightness~\cite{reviews}.

The problem of searching for neutrino mass and studying the oscillation 
phenomenon has been faced in the past through many different experimental 
techniques.
The first studies, based on the so called Fermi-Perrin method of the 
observation of the $\beta$ spectrum near the end point, gave 
\cite{primamassa} 
the limit $m_{\nu} \leq 500\ \rm MeV$.  
This limit was obviously lowered many times in the 
following years, up to the present results~\cite{nutau,numu,nue}.  

Since the experiment of Goldhaber et al. ('58)~\cite{Goldhaber}, we know 
that neutrinos produced in $\beta$ decays are left-handed particles.
This fact at the beginning appeared as a confirmation of the 
hypothesis that 
the neutrino is a massless particle.

Another milestone in the development of our knowledge of 
neutrino physics was
the idea suggested by Pontecorvo \cite{Pontecorvo} that neutrino can  
``oscillate'', in the sense that the flavor states are superposition of 
different mass states. This was a revolutionary hypothesis, because 
only the electronic neutrino was known in those days, but 
nowadays we have 
strong experimental hints that would 
confirm  the validity of Pontecorvo's 
idea. 

Coming to our days, in the usual version of the Standard Model that describes
the electroweak interactions, the neutrino is a left handed Dirac particle; 
hence, in such a theory it is impossible to build a renormalizable mass term 
for this particle. 
On the other hand, as we are going to see in detail, there is 
experimental evidence that it is a massive and oscillating particle.
Therefore we are forced by the data to enlarge this ``minimal version'' of the 
Standard Model and possibly to build a more general theory in which a neutrino
mass can be fitted naturally.

Neutrino physics can be considered an ideal playground to test different 
theories beyond the Standard Model, like, for instance, supersymmetry and 
grand unification theories~\cite{susyunified} or the theories based on the existence of large extradimensions \cite{extradimensions}.
The determination of the value of neutrino mass also has important 
implications on cosmological models. In particular neutrino is a candidate 
for dark matter and this fact determined a revival of astrophysical 
studies of neutrino properties in the last decade \cite{astroneutrini}.
 
\section{Evidences of neutrino mass and oscillations}
 
All the experiments aiming to measure the neutrino mass and to test the existence 
of oscillations can be classified in some main categories.

First of all there are the direct kinematical searches like the ones 
of~\cite{nutau,numu,nue} and the searches for the neutrinoless double 
$\beta$ decays~\cite{Klapdor-Kleingrothaus:yx,Aalseth:2002rf}.
The present limits on the values of $\nu_{\tau}$ and $\nu_{\mu}$ 
masses are~\cite{nutau,numu}: 
\begin{eqnarray}
m (\nu_{\tau}) & < & 18.2 \quad {\rm \MeV}\nonumber\\  
m (\nu_{\mu})  & < & 190 \quad {\rm keV} \, . 
\end{eqnarray}
The best limits for the mass of the electron neutrino, instead, have been obtained 
from the Mainz and the Troitsk \cite{Bonn:tw,Lobashev:uu} experiments which 
have found $m (\nu_{e})  < 2.2 \quad eV $. 
In future many experiments will try to lower this limit. In particular there 
is a great expectation for KATRIN
(the Karlsruhe Tritium Neutrino experiment) \cite{KATRIN}, that should 
start data taking in 2007 and improve the sensitivity down to 
$0.35 \, eV$.  

The search for neutrinoless double $\beta$ decays is 
important because the 
observation of these decays would be a clear indication in favor of a 
Majorana nature of the neutrino, if we assume CPT invariance~\cite{Pascoli:2001vr}.
The most stringent limit on this process available at the moment comes from 
the Heidelberg-Moskow collaboration~\cite{Klapdor-Kleingrothaus:yx} 
$ \langle m_{\nu} \rangle < 0.35 \, eV$ and 
from IGEX (International Germanium Experiment)~\cite{Aalseth:2002rf} 
$ \langle m_{\nu} \rangle < 0.33-1.35 \, eV$. 
In the last year there has been a claim \cite{Klapdor-Kleingrothaus:2001ke} 
from some members of the Heidelberg-Moskow collaboration of discovery of a 
$ 2.2 \sigma$ effect that would be a   
signal of neutrinoless double $\beta$ decay, but this result has been strongly 
contested and the discussion on its validity is still an open question.

A second group of experiments uses neutrino fluxes produced at accelerators 
and nuclear reactors. They are usually divided in long- and short-baseline,
according to the distance between the neutrino production point and 
the detector.

Many short baseline accelerator experiments didn't find any signal 
of oscillation.
They are nevertheless important, because they give constraints on 
the possible values of the mixing parameters. The most important limits have 
been obtained by NOMAD \cite{NOMAD} and CHORUS \cite{CHORUS} at CERN. 

These two experiments were designed to check relatively high values of 
the mass differences ($\Delta m^2 >_{\approx} 1 \, eV^2$) and used a beam of 
$\nu_{\mu}$ to look for a signal of a $\tau$ production, that would have been 
an indication of $\nu_{\mu} \to \nu_{\tau}$ oscillations .

Besides the long baseline reactor experiments it is worthwhile to recall 
the results of CHOOZ~\cite{CHOOZ} and Palo Verde~\cite{Wang:pn,Boehm:2001ik}. 
At CHOOZ a beam of reactor $\bar{\nu}_e$ was sent to a detector located 
about 1 Km away and detected through the reaction 
$\bar{\nu}_e +p \to e^+ + n$.  
No evidence of oscillation was found at CHOOZ and the experimental result 
for $R$, that is the ratio between the number of measured $\bar{\nu}_e$ events 
and the expected number in absence of oscillation, 
is compatible with $R = 1$. 
In a simple two flavor model the  oscillation probability is given by the 
relation 
\begin{equation}
P_{\nu_{\alpha} \to \nu_{\beta}}= \frac{1}{2} \sin^2 \theta
\left(1- \cos \left( 2.53 \frac{\Delta m^2 L}{E}\right) \right) \quad (\alpha \neq 
\beta),
\end{equation}
where $L$ is the distance source-detector expressed in meters, E is the $\nu$ energy in \MeV and $\Delta m^2$ is the difference of the squares of neutrino 
masses expressed in $\rm eV^2$. 
The range of mass differences and mixing angles that can be tested in a 
certain experiment is limited by the requirements that the source-detector distance 
be much shorter than the oscillation length 
$$\left(L_{osc}(m) \simeq 2.48 \frac{E({\rm \MeV})}{\Delta m^2(eV^2)}\right).$$
The CHOOZ average  energy value is $\langle E \rangle \approx 3 {\rm \MeV}$; 
therefore CHOOZ results can be used to exclude a significant part of the 
mixing parameters plane. In particular they tell us that $\Delta m^2$ must be 
smaller than $10^{-3} eV^2$, unless the values of the mixing angle are very 
small.  

The opposite situation took place in the case of LSND~\cite{Sung:ps,LSND}, a 
short baseline accelerator experiment performed with a neutrino beam produced 
at the Los Alamos meson physics facility (LAMPF).      
The experiment found evidence of two kinds of oscillation signals. The first 
was the excess of $\bar{\nu}_e$ in the beam of $\bar{\nu}_{\mu}$ produced by 
the decay at rest of the $\mu^+$ obtained as secondary products of the 
proton accelerator beam. The second was a signal of $\nu_{\mu} \to
\nu_e$ oscillations, starting from the $\nu_{\mu}$ produced by the $\pi^+$ decay in 
flight.
The LSND result, if confirmed, ought to be a clear indication of 
oscillation with very high values of the mass difference, up to $\Delta m^2 
\geq 1 eV^2$.
To reconciliate this result with the ones coming from solar and atmospheric 
neutrinos one would have to postulate the existence of at least one sterile 
neutrino in addition to the usual three active ones 
\footnote{for a recent discussion about the possible explanations 
of LSND data see~\cite{Strumia:2002fw}}.  
However, up to now there has been no independent confirmation of the LSND 
results.
The KARMEN experiment~\cite{Wolf:2001gu,karmen}, 
performed at the Rutherford 
Laboratories, explored a significant part of the mixing parameter space 
proposed by LSND and it didn't find any signal of oscillation 
\footnote{About the compatibility of LSND and Karmen results see 
also~\cite{LSNDconKARMEN}}. 
A new experiment MiniBoone~\cite{Hawker:pt,Stefanski:2001rk} is going to run 
very soon and produce data starting from 2004. It is very similar to LSND 
and will test definitely the validity of LSND results. 
     
A new generation of very long baseline experiments has become available in the 
last years. The forerunner of them is K2K~\cite{Yanagisawa:xz,Wilkes:2001vg,K2K}, that uses a neutrino beam produced at the Japan kaon facility KEK and 
detected at the Kamioka site. Up to now K2K has detected 56 events instead 
of the expected value in absence of oscillations of $80^{+7.3}_{-8}$ events. 
This is a confirmation of neutrino oscillations (the no-oscillation 
probability is less than $1 \%$). Moreover the best fit point~\cite{K2K} 
values for the mass difference  and the mixing angle 
($\Delta m^2 = 2.8 \times 10^{-3} eV^2$ and $\sin^2 2\theta = 1$) are in good 
agreement with the results of atmospheric neutrino experiments.
Two similar projects have already been approved and will become available 
in the near future: one of them is a   
neutrino beam from CERN to the Gran Sasso 
Labs~\cite{CERNGSasso,Weber:2001gw,Rico:2002mh,Arneodo:sg} 
and the other one is in the 
USA~\cite{Paolone:am,Lang:rw} (from FNAL to Soudan). 
The long baseline accelerator experiments will probably give
an important confirmation of the oscillation evidence which have up to now come from  
the study of solar and atmospheric neutrinos. They are also expected to 
find in an unambiguous way indications of oscillation from appearance signals. 
In addition, in the long baseline experiment one has the opportunity of 
choosing the specific characteristic of the beam; hence they can be used to 
perform precision measurements~\cite{Ota:2002na,Barger:2001yx}. 
For instance they should be useful to study 
the value of the mixing angle $\theta_{13}$, relevant for eventual CP 
violation. 
The present limit on the measurement of this angle coming from CHOOZ 
($\theta_{13} \leq 9$ degrees), could be lowered to the level of about 
$5$ degrees at ICARUS, one of the two experiments that will use the 
CERN-Gran Sasso beam.

Important results should very soon come from the long baseline reactor 
experiment KamLAND~\cite{kamLAND}, which might in principle give a definite 
solution to the solar neutrino problem, as we will see in the following.

The two main categories of experiments looking for oscillation signals are 
the ones that study the atmospheric and the solar neutrinos.
The atmospheric neutrinos are products of decay of the cosmic rays. The number
of electronic and muonic neutrinos can be computed with good accuracy, considering the properties of cosmic rays, their decay channels and eventually geomagnetic 
effects.
Most of the atmospheric neutrino experiments measure  the value of the double
ratio 
$$R= \frac{(\mu/e)_{data}}{(\mu/e)_{MC}}.$$
 The numerator and denominator 
 are respectively the experimental and the Monte Carlo computed values of 
the ratio between the events generated by muonic neutrinos (and antineutrinos)
and the ones generated by electronic neutrinos (antineutrinos).
There are essentially two kind of experiments: the water Cherenkov (like 
Kamiokande~\cite{Hatakeyama:1998ea,Oyama:bk,Fukuda:1994mc,Hirata:1992ku}, 
Super-Kamiokande~\cite{Fukuda:1998mi,Kajita:zv,Toshito:2001dk,Fukuda:2000np,SKatmres}, 
IMB~\cite{Becker-Szendy:vr}) and the iron plate calorimeters 
(like Soudan II~\cite{Sanchez:pj} and in the past years Frejus~\cite{Daum:bf,Berger:1989wy} 
and Nusex~\cite{Aglietta:1988be,Ragazzi:ti}). 
Clear evidence of oscillations has been found at Kamiokande, 
Super-Kamiokande (SK), IMB and Soudan II and also at the MACRO~\cite{Ambrosio:2001je,Giacomelli:2001td} experiment at Gran Sasso. 
The best statistic has been obtained at SK, which found~\cite{SKatmres}
$R = 0.638 \pm 0.016 \pm 0.050$ 
for the Sub-GeV events and 
$R = 0.658^{+0.030}_{-0.028} \pm 0.078$ for the Multi-GeV events.
Another interesting observable is the up-down asymmetry between the up 
going events, in which the neutrino crossed the Earth before interacting 
in the detector, and the down going ones: 
$A_{e,\mu}= (\frac{U-D}{U+D})$.   
The experimental value of this quantity is consistent with zero for the 
electronic neutrinos, while for the muonic ones the up-down asymmetry
for high values of the momenta is a decreasing negative value. 
These results are clear indications of a reduction of the flux of 
muonic neutrinos and antineurinos that arrive at the detector after crossing
the Earth. 
The most natural explanation of this phenomenon is the possibility that the 
muonic neutrinos oscillate into other flavors and the oscillation probability
is greatly enhanced by the interaction with matter.

The last group of experiments is that of the experiments observing the 
neutrinos coming from the Sun. We will discuss them in detail in the rest of 
the paper. 


\subsection{History of the solar neutrino problem}

The first experiment on solar neutrinos, Homestake~\cite{Homestake}, started at the end 
of the $`60s$ using the inverse $\beta$ decay on chlorine 
$^{37}Cl + \nu_e \to  ^{37}Ar + e^-$. The threshold energy was 
$E_{thr} \simeq 0.81 {\rm \MeV}$, hence it was sensitive to the pep, $^7 Be$, 
$^8 B$ and hep components of the solar neutrino flux.
The results were really surprising, because Homestake found a deficit of the solar
neutrino flux of more than $60 \%$ that predicted by the Solar 
Standard Model (SSM). The updated value of the ratio $R$, between the 
experimental results and the SSM prediction~\cite{BPB2000}, for the chlorine  
experiment is $R=0.34 \pm 0.03$.
This result raised fundamental questions: what happens to solar $\nu$ on their
 way to earth? Eventually, could the SSM be wrong? 

The Homestake indication was confirmed by similar experiments, SAGE~\cite{SAGE}
in Russia and GALLEX~\cite{GALLEX} and later on GNO~\cite{GNO} at the INFN 
Gran Sasso Labs, which used gallium instead of chlorine.
The energy threshold is lower in the gallium experiments 
($E_{thr} \simeq 233 \, keV$) making them also sensitive to pp
neutrinos, which are the main component of the solar neutrino flux.  
The updated gallium results are~\cite{SAGE,GALLEX,GNO}:
\begin{eqnarray}
R & = & 0.60 \pm 0.05 \quad (SAGE) \nonumber\\  
R & = & 0.58 \pm 0.05 \quad (GALLEX-GNO) \, . 
\end{eqnarray}
This confirmation of Homestake results gave a strong support to the neutrino 
oscillation hypothesis and caused an increase of the interest for this problem.
It could be a signal of {\it new physics}.

An essential improvement in the knowledge of solar neutrinos came with the 
advent of the water Cherenkov experiments, Kamiokande~\cite{Kamiokande} and
Super-Kamiokande (SK)~\cite{SKsolar,Smy:2002hr}, that 
looked at the elastic scattering $\nu_e + e^- \to \nu_e + e^-$ and 
confirmed the existence of the ``solar neutrino problem'' with a very high 
statistic. In this experiments it was possible to know the direction of the 
incoming neutrino (by looking at the outgoing direction of the recoil electron)
and also to study the energy and angular spectrum and the day night 
asymmetries.  
The energy threshold for these experiments was quite high (5 \MeV for SK) 
and therefore they were sensitive only to the high energy component of the neutrino flux, that is $^8 B$ and hep neutrinos. Their results confirmed the existence of a deficit in the electron neutrinos reaching the detector. 
The SK result for the energy spectrum and the small values of the day-night 
asymmetries were also very important to put strong constraints on the possible
 values of the mixing parameters.  

After the publication of SK data it was clear that there was a deficit 
of solar electron neutrinos reaching the Earth, with respect to the flux predicted by SSM. The oscillation hypothesis was considered the most 
plausible explanation of this phenomenon, but there were still different 
regions allowed by the experiments in the mixing parameter plane, as we will 
see in detail. 

\subsection{The post SNO situation}

The real breakthrough was due to the SNO experiment that published its first 
data in 2001~\cite{SNOCC}. 
SNO is a deuterium Cherenkov detector designed to 
look simultaneously at three different reactions:
\begin{equation}
\begin{array}{lllll}
\nu_e + d & \to & e^- + p +p &  \text{{\sl (Charged Current)};}\\
\nu_x + d & \to & e^- + n +p & \text{{\sl (Neutral Current)};}\\  
\nu_x + e^- & \to & \nu_x +e^-  & \text{{\sl (Elastic Scattering)}.} 
\end{array}
\end{equation}
The first reaction (CC) receives contribution only from the electron neutrino,
while the others (NC and ES) are sensitive to all neutrino flavors.
This experiment gives the first direct model independent measurement of the 
total solar neutrino flux reaching the Earth (through the NC observation) and 
at the same time, comparing this flux with the one of $\nu_e$ recovered from 
CC, it offers a strong evidence of the oscillation of $\nu_e$ into other 
active neutrinos.

During its first phase of working~\cite{SNOCC} SNO observed the charged current
and elastic scattering events, with an energy threshold for electron
detection of 6.75 \MeV.
The $\nu_e$ flux measured from CC, after 241 days of running, was: 
$\Phi_{\nu_e}^{CC} = 1.75 \pm 0.07 (stat.)^{+0.12}_{-0.11} (syst.) \times 10^6 
cm^{-2} s^{-1}$. The ratio between this value and the SSM prediction was 
$R=0.35 \pm 0.03$. 
In the SNO experiment the neutrino flux can be recovered also from the elastic
scattering, using the relation \footnote{the contribution of $\nu_{\mu}$ 
and $\nu_{\tau}$ to the elastic scattering cross section is only through 
neutral current, hence it is about 1/6 of the contribution of $\nu_e$ that can 
interact also through charged current}:
\begin{equation}
\Phi_{\nu}^{ES} = \Phi_{\nu_e}^{ES} +0.154 \sum_{i=\mu,\tau} 
\Phi_{\nu_i}^{ES}.
\end{equation} 
The value of the total neutrino flux recovered from the elastic scattering at 
SNO and also, with a better statistics, at SK doesn't agree with the $\nu_e$
flux obtained from SNO CC. The comparison of the two results gives:
\begin{equation}
\sum_{i=\mu,\tau} \Phi_{\nu_i}^{ES} = 3.69 \pm 1.13 \times 10^6 cm^{-2} 
s^{-1}. 
\end{equation}
This result was the first evidence (at $3 \sigma$ level) of the 
presence of muonic and tauonic neutrinos in a electronic neutrino beam 
reaching the Earth from the Sun. Therefore it was, up to the present SNO
data on NC, the most robust evidence of $\nu_e$ oscillation into other active 
neutrinos.
It is also remarkable that the sum of the $\nu_e$ and $\nu_{\mu,\tau}$ 
fluxes give a value in good agreement with the SSM prediction. Consequently 
the results of SNO phase I also strongly disfavored the 
hypothesis of pure oscillation into sterile neutrinos.

Recently, while this paper was in preparation, the data of the so called
phase II of SNO also became available~\cite{Ahmad:2002jz,Ahmad:2002ka}. 
This data, obtained with 306.4 days of 
running, confirms the indications of the phase I and includes the first 
neutral currents (NC) observations. 

\section{Global analysis of the solar neutrino data}

Given all the experimental data that we have just reported, one can 
say that there is really strong evidence that neutrinos are massive 
and oscillating particles. Nevertheless, many details of the 
mass patterns still have to be clarified.
With this aim in mind, we developed a global analysis of all the available data  
on solar neutrinos, also including the CHOOZ constraints.  
Our first purpose was that of determining the regions in the mixing parameter 
plane that are still compatible with the experiments.
In addition to this, we wanted to understand how the forthcoming experiments
(in particular Borexino and KamLAND) could improve our knowledge of neutrino 
mass properties.

We assumed neutrino oscillation as a working hypothesis and considered 
bidimensional models. 
For details of our analysis we refer the interested reader 
to~\cite{Aliani:2001zi,Aliani:2002ma}.
Here we just report the most salient aspects of our strategy.
The analysis is based on the numerical calculation of the expected 
event rate for every solar neutrino experiment as a function of the mixing 
parameters and on the comparison between these expected numbers and the 
experimental data.
The statistical analysis is based on the $\chi^2$ method its output being
 contour plots in which one can see which values of the mixing angles
and mass differences are still allowed at a given confidence level.

Our calculation can  essentially be split into two parts. The first one is 
the determination of the neutrino transition amplitude, i.e. the 
probability for an electronic neutrino produced in the Sun to change its 
flavor before reaching the detector.  
The other ingredient is the calculation of the detector response functions, 
that, for a given neutrino energy, depend on the experimental details of the 
specific detector (i.e. efficiency, resolution, etc.) and on the cross 
section for the reaction under examination.

The transition amplitude calculation is separated in three parts, 
corresponding to the neutrino propagation inside the Sun, in the vacuum and in the Earth. For every value of the mixing parameters we compute fully 
numerically the amplitudes in the Sun and in the Earth, while the one corresponding to the vacuum evolution is computed analytically. The three amplitudes are patched together using the evolution operator formalism
\cite{Torrente-Lujan:1998sy}. 

\section{The present situation}

We included in our analysis the total rates of the chlorine and gallium 
experiments, together with the different energy bins of SuperKamiokande and  
with the charged current results of the first phase of SNO.
The resulting contour plots are reported in Figure 1.
One can note that there are still two allowed region, even at 
$90 \%$ C.L.: 
the 
Large Mixing Angle (LMA), where we found the best fit point with a goodness of
fit (g.o.f.) of $84.38 \%$, and the so called LOW solution, characterized by 
lower values of the mass differences.
The two other possible solutions of the solar neutrino problem, that 
historically have been considered, are the Small Mixing Angle (SMA) and the 
vacuum 
solution, corresponding to much lower values of $\Delta m^2$.
These two solutions cannot be completely excluded, even if they are strongly 
disfavored mainly by SK data on the day and night energy spectrum and by the 
SNO results.
Another effect of SNO data was that of shrinking the different regions which 
became well separated.
Our results are in good agreement with most of the analysis one could find in
 literature~\cite{Bahcall:2001zu}.


\begin{figure}[h]
\centering
\includegraphics[scale=0.6]{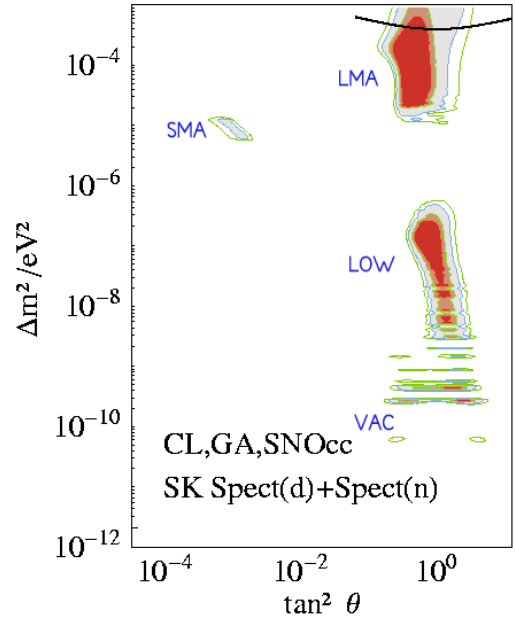}
\caption{\it The situation after the charged current SNO data. The different colored regions correspond to different confidence levels: $90\%, 95\%, 99 \%$ and  $99.7 \%$.}
\label{specfig}
\end{figure} 

\section{Future scenarios}

Given this situation, we studied which new information should come in future 
from the Borexino data~\cite{Aliani:2001ba,Aliani:2002rv}. 
Borexino~\cite{Borexino,Meroni:zj} is a solar neutrino experiment, mainly sensitive to 
the $^7 Be$ component of the neutrino flux, that should start running in very 
next years at the Gran Sasso Labs. In Figure 2 the usual contour plots 
obtained from all the experiments available up to now are superimposed to the 
contour lines corresponding to different hypothetical possible values
 of the total rate at Borexino.
As one can see from the picture, Borexino should be able to clarify  
the situation in the case in which the solution  very well is in the small mixing angle region. The situation would be, instead, more complicate in case of LMA or LOW 
solutions. In these two regions, in fact, the ratio between the Borexino 
signal and the SSM prediction in absence of oscillations should be between 
0.6 and 0.7 .
The discrimination power of Borexino increases a lot if we look also at the 
day-night asymmetry, as one can see from Figure 3.
The LOW region is characterized by high values of the asymmetry, that can reach
up to $ 20 \%$, while in the LMA region the day-night asymmetry is much lower.
Hence, by looking simultaneously at the total rate and at the day-night 
asymmetry, Borexino should be able to discriminate between the two solutions 
of the solar neutrino problem that are compatible with the experiments up to 
now, that is the LMA and the LOW solutions.   

\begin{figure}
\centering
\includegraphics[scale=0.6]{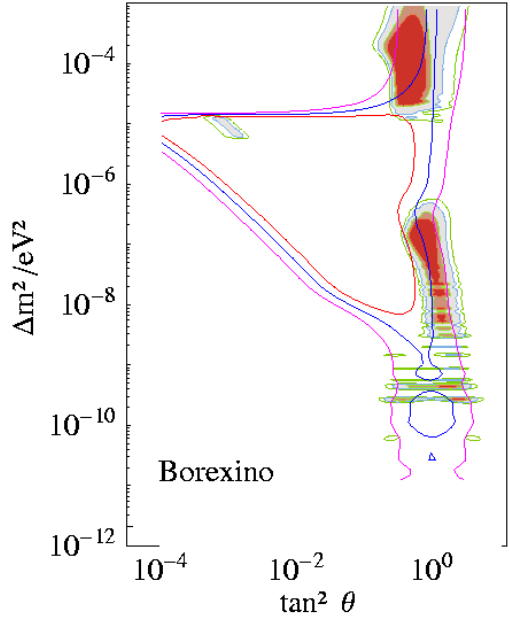}
\caption{\it
Possible values of the total signal at Borexino. The full lines correspond to a total rate equal respectively to 0.5, 0.6 and 0.7 with respect to the SSM 
prediction in absence of oscillation. These isosignal lines are superimposed   to the contour plots (colored regions) corresponding to the regions allowed 
at different confidence levels from the other experiments.}
\label{borexinototal}
\end{figure} 

\begin{figure}
\centering
\includegraphics[scale=0.6]{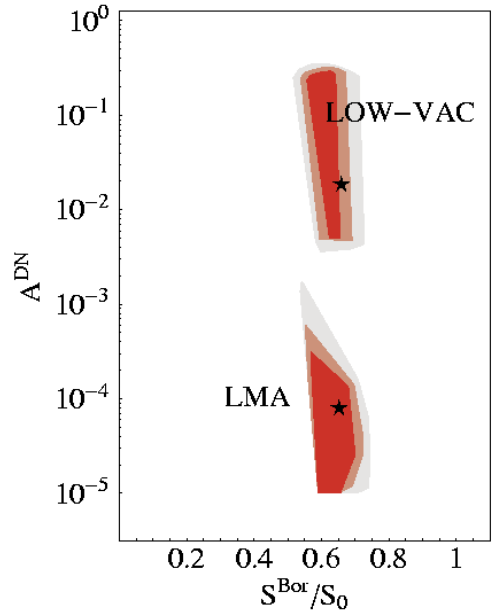}
\caption{\it Predicted values of the day-night asymmetry at Borexino.}
\label{Borexinodaynight}
\end{figure} 

Another very important experiment, already running, that should significantly improve
our knowledge of the mixing parameters relevant for solar 
neutrinos is KamLAND~\cite{kamLAND}. 
In this experiment a flux of low energy $\bar{\nu}_e$ produced by different nuclear reactors is sent to a scintillator detector capable of detecting their interactions with protons.  
Although it is not a traditional solar
neutrino experiment, KamLAND is sensitive to neutrino oscillations with 
mixing parameters in the LMA region, that seems to be the solution of the 
solar neutrino problem preferred by the present data.
Therefore, we can hope that KamLAND will soon be able to determine the exact
values of the mixing parameters with satisfactory accuracy. 
The main limitation of KamLAND is its reduced sensitivity to the extreme 
upper part of the LMA region, that could create problems in the determination 
of $\Delta m_{12}^2$,  as discussed in~\cite{Petcov:2001sy} and later on 
in~\cite{HLMA}. 
For a detailed discussion about KamLAND potentiality and discrimination power 
we refer the interested reader to~\cite{Aliani:2002rv}.

While this paper was in preparation the data of the second phase of SNO 
was published~\cite{Ahmad:2002jz,Ahmad:2002ka}.
They contained the first direct oservation of the neutral current (NC) process
 and the data of the CC and ES processes with statistics higher than the one the first 
SNO phase~\cite{SNOCC}.
From the NC data one can recover a value of the total active flux 
$\Phi^{NC}_{^8 B} (\nu_{TOT}) = 5.09^{+0.44}_{-0.43} (syst.) ^{+0.46}_{-0.43}
(stat.)$ which is in very good agreement with the SSM prediction 
( $\Phi_{^8 B} (\nu_{e}) = 5.05 \times 10^6 cm^{-2} s^{-1}$,
if we use the old value for $S_{17}$). This result is an important confirmation of the validity of the SSM. At the same time, comparing these values 
with the SK and SNO ES values of the $^8 B$ electron neutrino flux 
one obtains proof that a significant part of the electron neutrinos coming 
from the Sun is converted into other active flavors 
(there is a $5.3 \sigma$ evidence that the flux of $\nu_{\mu,\tau}$ is 
different from zero).

We have redone our analysis with the addition of these recent SNO data.
In~\cite{Aliani:2002ma} we have assumed the simplifying hypothesis that the 
spectrum is undistorted with respect to the form predicted by the SSM in 
absence of oscillations. This assumption is essentially valid in the LMA
region (the one preferred by the data at the moment).
For the detailed values of the mixing parameters we recovered in this region 
and for the related study of KamLAND potentiality we refer the reader 
to~\cite{Aliani:2002ma}. Here we just recall that our results are in good 
agreement with other similar analysis~\cite{Bahcall:2002hv}.
We are also doing a more sophisticated analysis of the full mixing parameter plane, without the undistorted spectrum hypothesis and any other model dependent assumption.

A critical analysis of the influence of the different experimental results
and of the possible experiments that should come after Borexino and KamLAND is
performed, for instance, in~\cite{Strumia:2001gi}    

\section{Acknowledgements}
One of us V.A. would like to thank the organizers of the Conference 
``Les Rencontres de Physique de la Vallee d'Aoste'' for the invitation 
and the organizers and participants of the Conference for the nice and 
stimulating atmosphere. We all are grateful to Prof. S. Petcov for useful
and interesting discussions.
\end{document}